\documentclass[preprint,12pt,a4paper]{elsarticle}

\usepackage{amssymb}
\usepackage{amsmath,amsfonts}
\usepackage{xspace}
\usepackage{upgreek}
\usepackage{hyperref}

\usepackage{lineno}

\linenumbers

\biboptions{comma,square,numbers,sort&compress}
\newcommand{\pt}{\ensuremath{p_{\rm T}}\xspace}
\newcommand{\pT}{\ensuremath{p_{\rm T}}\xspace}

\newcommand{\gevc}[1]{\ensuremath{#1 \text{ GeV/$c$}}\xspace}

\newcommand{\snnt}[1]{\ensuremath{\sqrt{s_{NN}} = #1 \text{ TeV}}\xspace}

\newcommand{\pp}{pp\xspace}
\newcommand{\ppb}{p-Pb\xspace}
\newcommand{\pbpb}{Pb-Pb\xspace}
\newcommand{\NN}{A-A\xspace}

\newcommand{\LA}{\ensuremath{\Lambda}\xspace}
\newcommand{\AL}{\ensuremath{\bar{\Lambda}}\xspace}
\newcommand{\LQCD}{\ensuremath{\Lambda_{\rm QCD}}\xspace}
\newcommand{\KOs}{\ensuremath{{\rm K}^0_s}\xspace}

\newcommand{\dndeta}{\ensuremath{{\rm d}N_{\rm ch}/{\rm d}\eta}\xspace}

\journal{Nuclear Physics A}

\begin{document}

\begin{frontmatter}



\title{Identified Particle Production in p-Pb Collisions Measured with the ALICE Detector}


\author[label1]{Peter Christiansen}

\address[label1]{Lund University, Division of Particle Physics, Sweden}

\author{for the ALICE Collaboration}

\begin{abstract}
ALICE has unique capabilities among the LHC experiments for particle
identification (PID) at mid-rapidity ($|\eta| < 0.9$) over a wide range of
transverse momentum (\pt). In this proceeding~\footnote{For the International
  Conference on the Initial Stages in High-Energy Nuclear Collisions
  (IS2013).} recent measurements of \pT spectra for ${ \pi}$, K, \KOs, p, and
\LA in \ppb collisions are presented and compared to results from \pbpb and
\pp. In particular the implications for the question of the existence of
radial flow in small systems is discussed.
\end{abstract}


\end{frontmatter}


\section{Introduction}
\label{}

The measurement of transverse momentum spectra is of fundamental interest in
hadronic collisions as it provides insight into a wide variety of QCD
physics. At low transverse momentum, \pt, where perturbative QCD is not
applicable the spectra have to be modeled using phenomenological
approaches. In central \pbpb collisions the spectra of light flavor hadrons,
which is the focus here: ${\rm \pi}$, K, \KOs, p, and \LA, have been shown to be
successfully described using models relying primarily on the formation of a
medium that expands following nearly ideal hydrodynamics and hadronizes
according to the statistical thermal model~\cite{Abelev:2012wca}. However, in
general the same models have been found to provide a poor description of the
measured spectra in peripheral collisions~\cite{Abelev:2013vea}. The failure
of the simple combination of hydrodynamics and thermal spectra in peripheral
collisions suggests that the spectra there are closer to \pp like QCD
spectra. This is supportive of the basic idea that a Quark Gluon Plasma (QGP)
medium is produced in central collisions, but not in peripheral
collisions~\footnote{See G.~Roland these proceedings for a much more detailed
  account of what we have experimentally learned so far on the complicated
  relation between \pp, \ppb, and \pbpb.}.

However, it has recently been discovered at LHC that in \ppb collisions a
double ridge structure reminiscent of azimuthal flow is
observed~\cite{Abelev:2012ola,CMS:2012qk,Aad:2012gla}, and a similar structure
has been observed reanalyzing the existing RHIC
data~\cite{Adare:2013piz}~\footnote{See also R.~Venogupolan, A.~Sickles,
  F.~Wang, J.~Velskova, P.~Bozek, and P.~Kuijer in these proceedings}. To
follow up on this observation ALICE has published light flavor hadron spectra
as a function of multiplicity in \ppb~\cite{Abelev:2013haa}. As the data is
already published the goal here is only to discuss some aspects of the
results~\footnote{See also F.~Barile in these proceedings.}.

\section{Model Driven Interpretation}
\label{sec:model}

\begin{figure}[htbp]
  \centering
  \includegraphics[keepaspectratio, width=0.8\columnwidth]{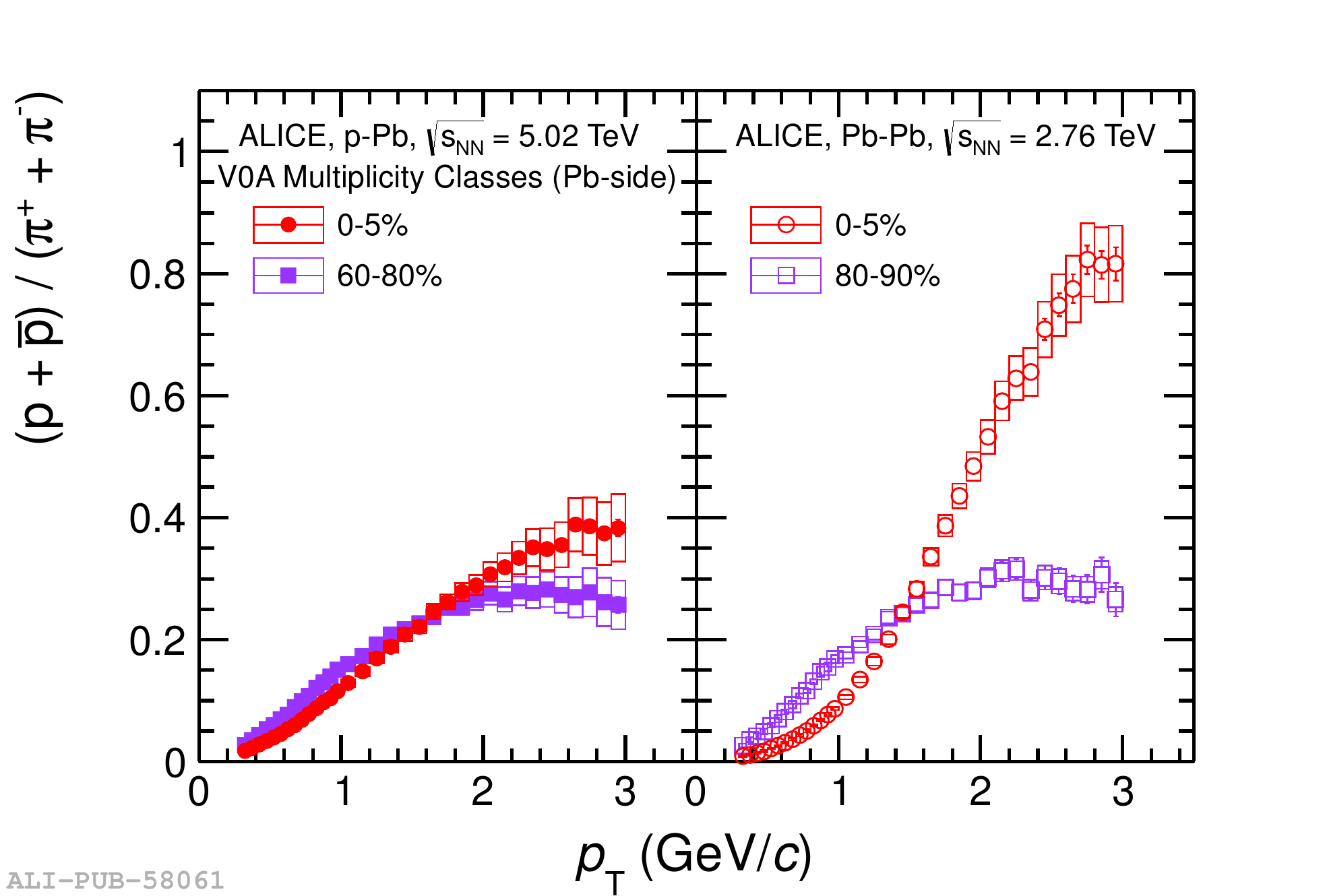}
  \caption{The proton-to-pion ratio, $(\rm p+\bar{p})/({\rm \pi}^{+}+{\rm \pi}^{-})$,
    as a function of \pt in \pp collisions for \ppb at \snnt{5.02} in two
    multiplicity classes (left) and for \pbpb at \snnt{2.76} in two centrality
    classes (right). Total systematic uncertainties are shown as open boxes
    and for \ppb results the uncorrelated systematic uncertainty is shown by
    the (small) solid boxes.}
  \label{fig:ratios}
\end{figure}

Figure~\ref{fig:ratios} shows the essential observation that will be discussed
in this proceeding; when the proton-to-pion ratio, $(\rm
p+\bar{p})/({\rm \pi}^{+}+{\rm \pi}^{-})$, in a high multiplicity class is compared to
a lower multiplicity class in \ppb~\footnote{A discussion of \ppb multiplicity
  and its relation to centrality can be found in the contribution from A.~Toia
  to these proceedings.}, we observe the same phenomena as for \pbpb
collisions: a decrease at low \pT and an increase at high \pT so that it
suggests that the protons have been ``pushed'' out to higher \pT. Here it is
important to stress that for the \ppb analysis the uncorrelated systematic
uncertainty for different multiplicity classes was evaluated and found to be
much smaller than the total systematic uncertainty, so the relative
multiplicity dependence of the proton-to-pion ratio is known with great
precision while the absolute ratio has significant systematic uncertainty.

\begin{figure}[htbp]
  \centering
  \includegraphics[keepaspectratio, width=0.49\columnwidth]{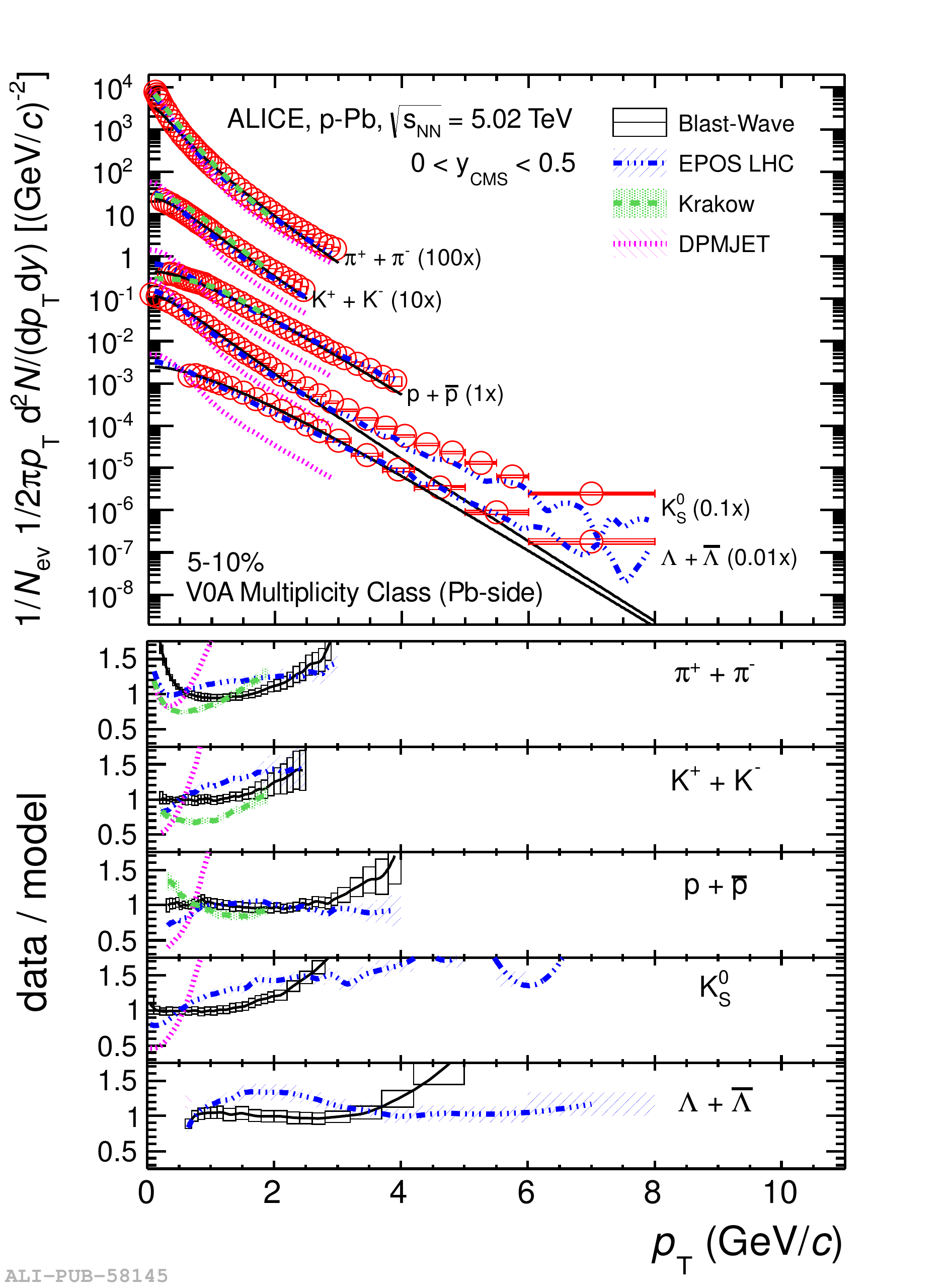}\\
  \includegraphics[keepaspectratio, width=0.49\columnwidth]{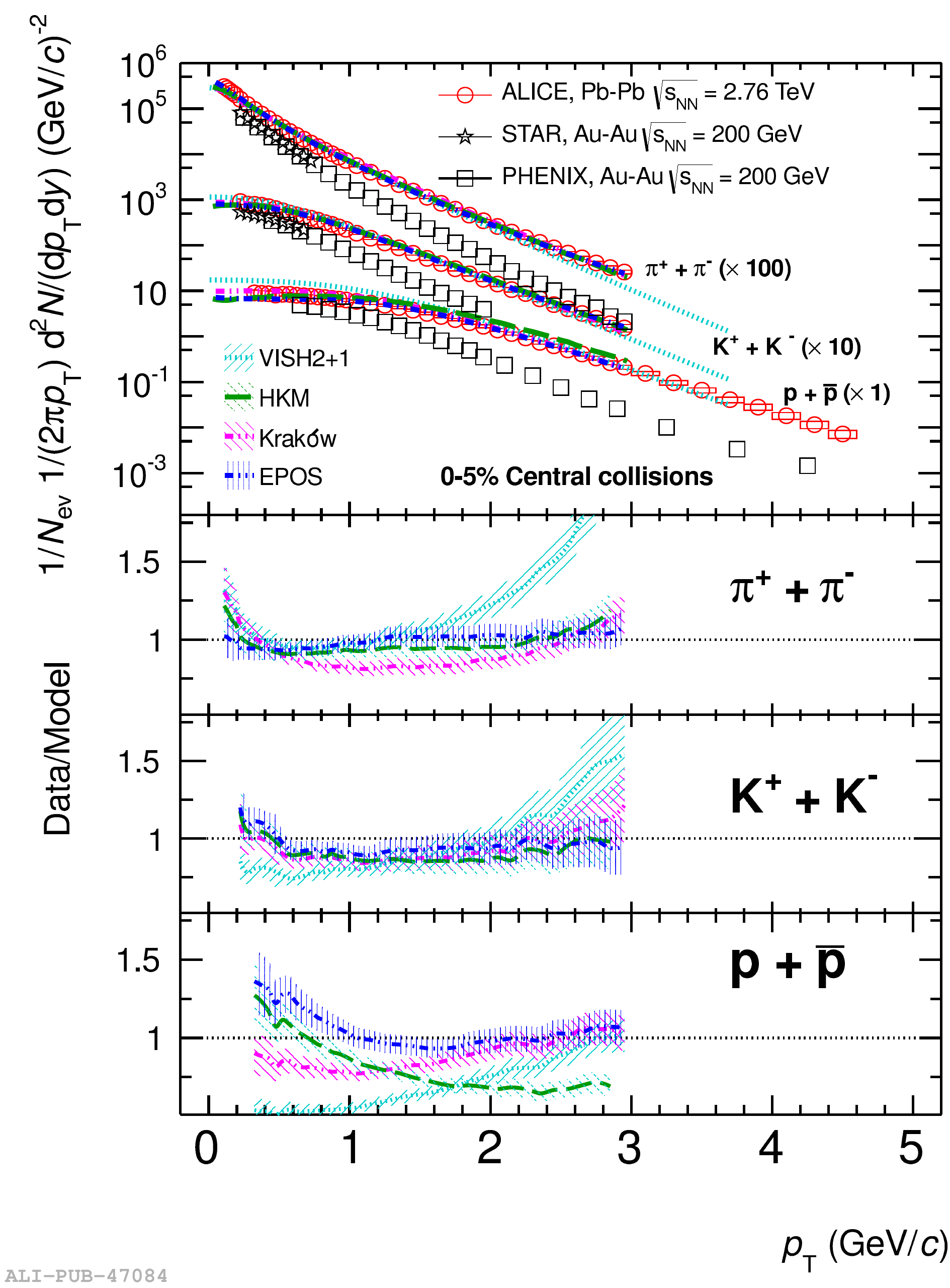}
  \includegraphics[keepaspectratio, width=0.49\columnwidth]{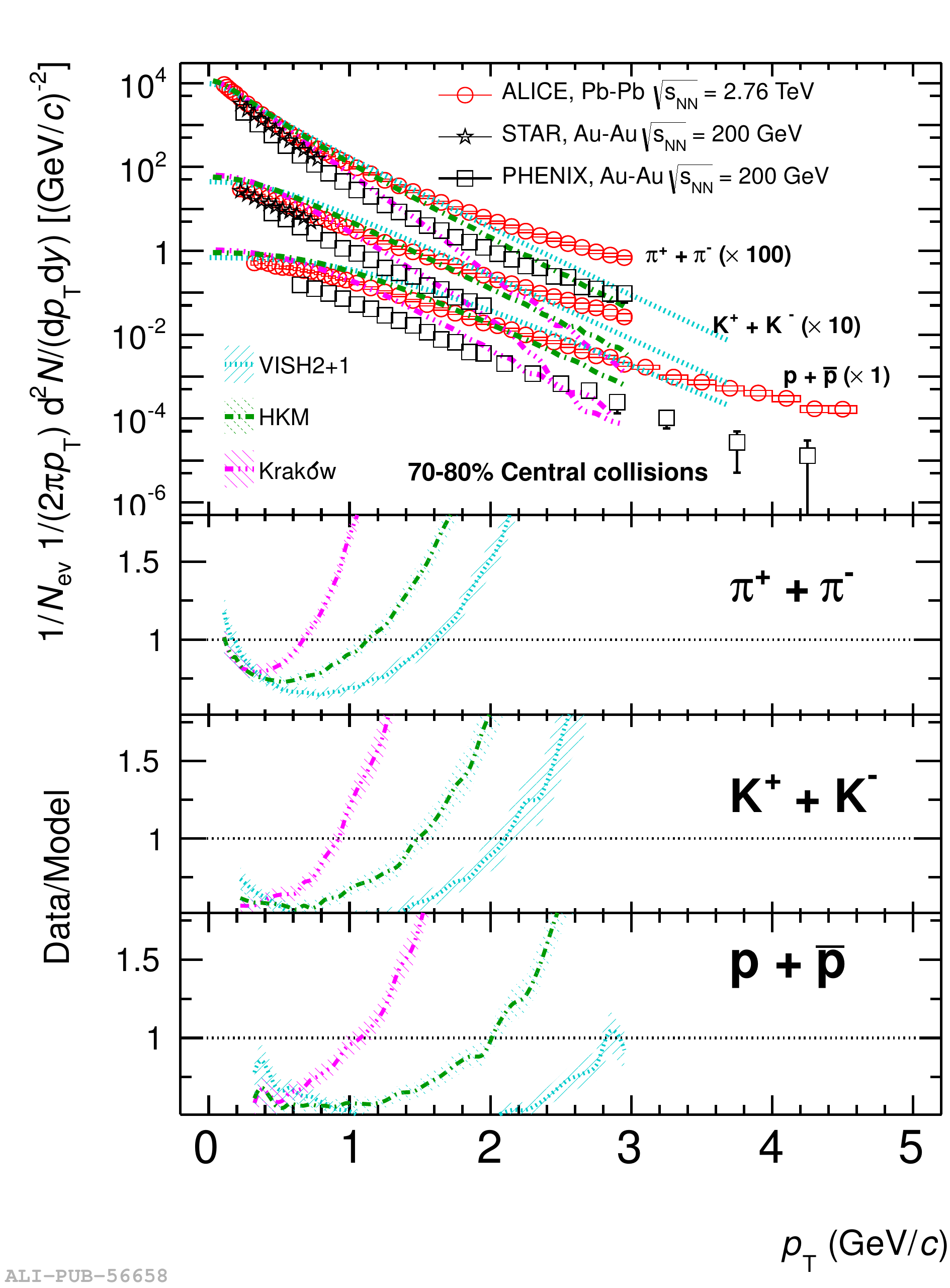}
  \caption{Light hadron spectra compared to various models for high
    multiplicity \ppb (top) and \pbpb central (bottom left) and peripheral
    (bottom right) collisions. All models, except for DPMJET (top plot only),
    incorporate a hydrodynamic phase, but have various implementation of
    additional physics in the hadronic phase.}
  \label{fig:spectra_vs_models}
\end{figure}

This behavior is reminiscent of radial flow where the heavier particles are
pushed out to higher \pT by the collective flow velocity
boost. Figure~\ref{fig:spectra_vs_models} shows the measured \pT spectra for
${\rm \pi}$, K, p, (\ppb and \pbpb) and \KOs and \LA (\ppb only) compared to
various models that, except for DPMJET (top plot only), all incorporates a
hydrodynamic phase. For a detailed discussion we refer to the published
references~\cite{Abelev:2013haa,Abelev:2013vea,Abelev:2012wca} and references
therein. One notes that except for peripheral \pbpb collisions in general the
hydro based models give a good qualitative and in some cases quantitative
description of the data. DPMJET which is based on PHOJET extended to nuclear
events via Glauber-Gribov theory fails to describe the data suggesting that a
hydro-like push is needed to describe the \pT spectra for high multiplicity
\ppb collisions. It is important to note that in the bottom right plot also
the relative trends for ${\rm \pi}$, K, p are different so that it is not just an
overall softening or hardening that is missing in the model description.

The main point that the author wishes to point out here is that while the
hydro approach fails in peripheral \pbpb collisions it works quite
successfully for p-Pb -- a system of a comparable size. If hydrodynamics is at
work in such small systems, as suggested by \ppb results, then something else,
like the understanding of the initial state in nuclear collisions, is needed
to explain why it fails for peripheral collisions.

Further studies can be found in the above paper where hydro-inspired blast
wave fits are also used to analyze \ppb and \pbpb data~\cite{Abelev:2013haa}.

\section{Data Driven Interpretation}
\label{sec:data}

\begin{figure}[htbp]
  \centering
  \includegraphics[keepaspectratio, width=0.49\columnwidth]{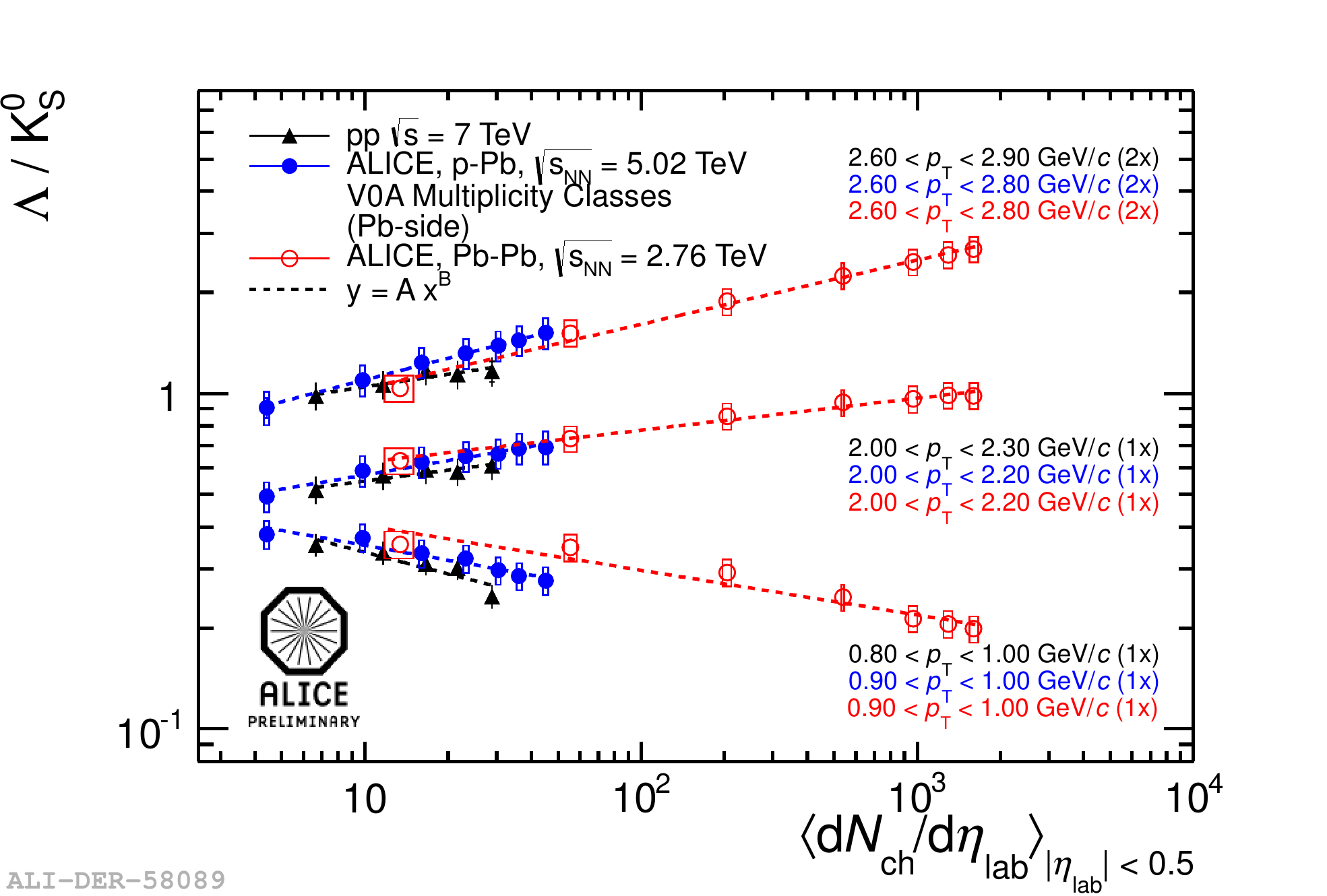}
  \includegraphics[keepaspectratio, width=0.49\columnwidth]{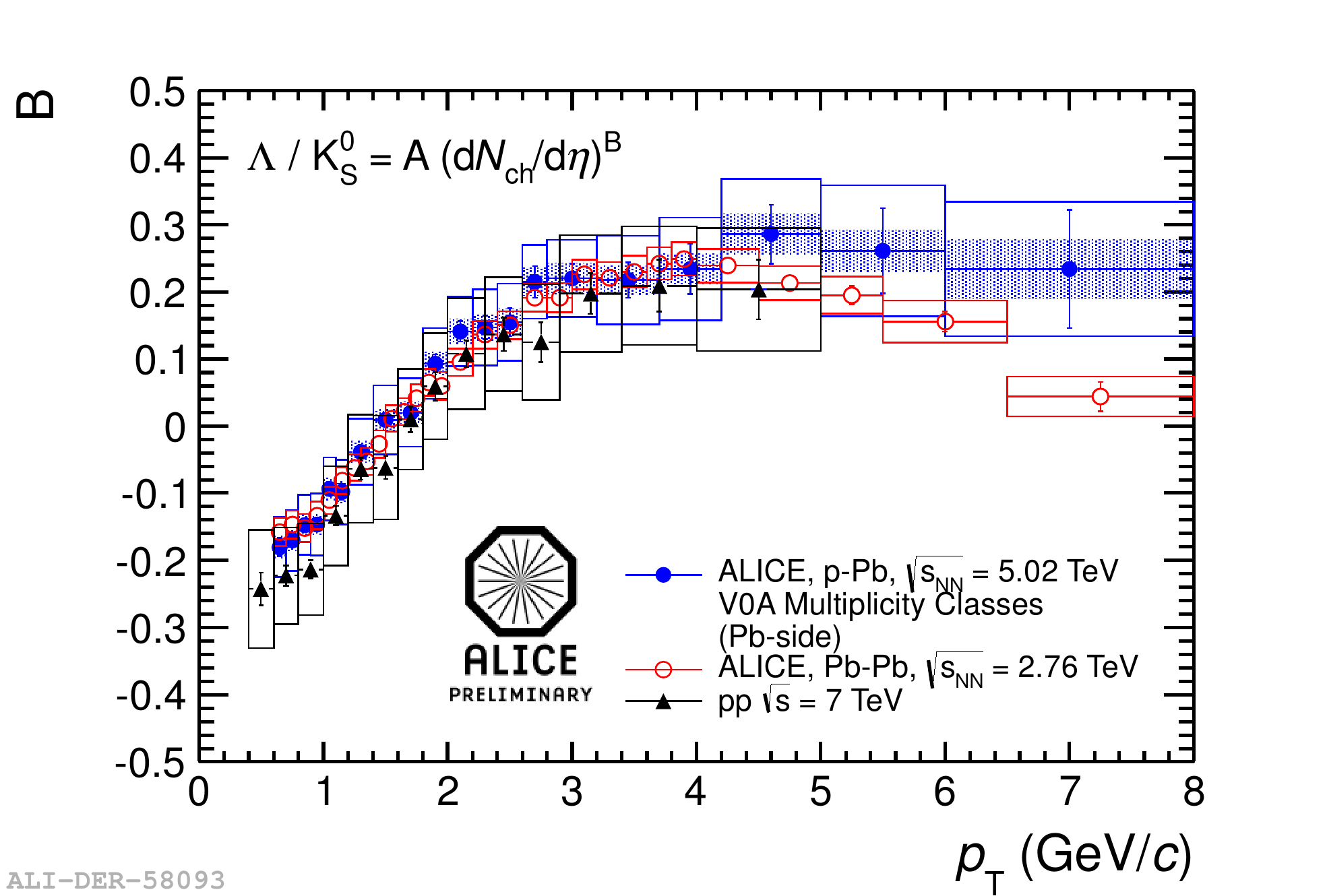}
  \caption{Left: the lambda-to-kaon ratio, $\LA/\KOs$, as a function of the
    mid-rapidity \dndeta in three \pt intervals for all measured multiplicity
    classes (\pp and \ppb) and centrality classes (\pbpb). The dashed lines
    shows power law fits for individual colliding systems. Right: the \pT
    dependence of the power law exponent from the fits for the 3 systems. As
    can be seen the relative increase of the ratio with \dndeta is similar in
    all cases even out to large \pT.}
  \label{fig:fits}
\end{figure}

Figure~\ref{fig:fits} shows an alternative data driven method of summarizing
the results. In the left figure the \LA/\KOs ratio is shown as a function of
the mid-rapidity \dndeta in 3 different \pT bins for all \pbpb centrality
classes and for all \pp and \ppb multiplicity classes. For each \pT bin and
system, \pp, \ppb, and \pbpb, the data points are fitted with a power law
function: $\LA/\KOs = A \cdot (\dndeta)^B$. The right figure shows the
exponent $B$ vs \pT for all 3 systems and it is evident that in fact this
exponent, within statistical and systematic uncertainties, is the same for all
systems. This is a surprisingly suggestive pattern that similar physics is
driving the ratios for all systems and for all sizes (no evidence for an onset
in \dndeta).

Is scaling of the particle ratios with the \dndeta intuitive? It does not
appear to be so. In a hydro picture the initial spatial geometry must be very
important while in this scaling relation it does not enter. But this might
not be all bad, as we just saw in the discussion of
figure~\ref{fig:spectra_vs_models} it is not possible for hydro based models
to describe both central and peripheral data, so maybe the \dndeta scaling is
more fundamental.

\begin{figure}[htbp]
  \centering
  \includegraphics[keepaspectratio, width=0.8\columnwidth]{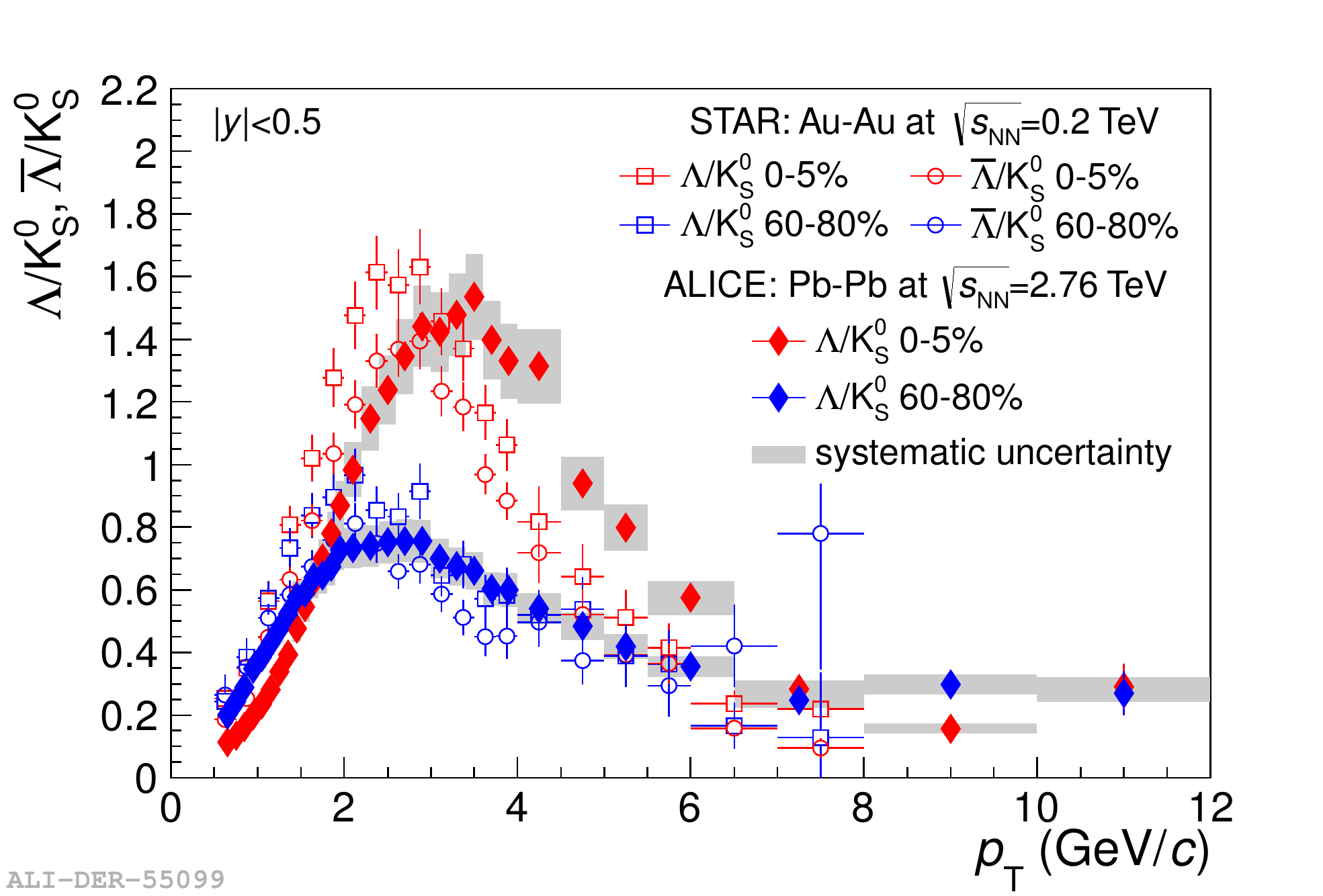}
  \caption{the lambda-to-kaon ratio, $\LA/\KOs$, as a function of \pT for
    central and peripheral \NN collisions at \snnt{0.2} (RHIC) and \snnt{2.76}
    (LHC).}
  \label{fig:LK_ratios}
\end{figure}

Another way to vary \dndeta is by varying the beam energy. Comparing LHC
(\snnt{2.76}) and RHIC (\snnt{0.2}) \dndeta at mid-rapidity it was found that
the increase for all centrality classes was approximately a factor
2.1~\cite{Aamodt:2010cz}. So one could expect that if the \dndeta scaling was
fundamental then the evolution from peripheral to central collisions should be
the same at both energies. Figure~\ref{fig:LK_ratios} shows a comparison
between the $\LA/\KOs$ ratios at LHC and RHIC for central and peripheral
events. As can be seen the results for peripheral events are quite similar
when the difference in baryon chemical potential, $\mu_{\rm B}$, at RHIC and
LHC is taken into account~\footnote{In statistical thermal models one would
  expect $(\LA + \AL)/\KOs$ to be approximately independent of baryon chemical
  potential when $\mu_{\rm B} \ll T$.}. However, for central events the
``push'' in \pT at LHC is significantly larger (as expected from hydro
calculations). It would be interesting to extend
figure~\ref{fig:spectra_vs_models} with similar comparisons for RHIC data
and the same models. A naive interpretation of the comparison is that it
agrees better with the traditional peripheral/QCD and central/QGP picture.

\section{Conclusion and Outlook}
\label{sec:concl}

Some of the ALICE results for production of ${\rm \pi}$, K, \KOs, p, and \LA have
been presented and discussed. In particular the similarity between the
different collision systems, \pp, \ppb, and \pbpb, have been pointed out,
where the lack of an onset in system size is surprising and in contrast to
ideas of rare fluctuations in small systems~\footnote{See, e.g., the
  discussion about ``fat protons'' by B.~Mueller in these proceedings}. We
have also tried to understand how the model predictions relate across system
sizes and beam energies to see if a simple model can describe all these
systems, and found that this does not seem to be the case.

Another question that is important for the outlook is how the small QCD
systems can have collective degrees of freedom. In QCD for light quark systems
there is no simple relation between the hadronic states and the quarks, e.g.,
for the masses, we rather need the scale $\LQCD \approx 200~\text{MeV}$. With
this scale we can get a rough feeling for the hadronic sizes, $r \sim \hbar
c/\LQCD$ and masses. As hadrons are made of 2 or 3 valence quarks we have a
good basis to assume that this corresponds to ``a unit of QCD
matter''. Typically the idea has been that, as in condensed matter physics
when grouping atoms together, there are emergent features when we group many
units of QCD matter together. These emergent features are supposedly the main
reasons to study the Quark Gluon Plasma. So by introducing hydrodynamics in
small systems the feature becomes QCD like rather than QGP like.

To make progress on the underlying physics one can look for inspiration also
in \pp models. PYTHIA contains two interesting ingredients.\\ On one hand it
is found that Multi Parton Interactions do not hadronize independently but
need a new ingredient where the solution so far has been color reconnection
which has recently been shown to give radial flow-like
boosts~\cite{Ortiz:2013yxa}. Color reconnection has no unique implementation
and there is an interest in \pp physics to explore alternative
ideas~\cite{Sjostrand:2013cya}. \\ On the other hand it is also known that to
remove the divergence of the hard cross section at low \pT one needs a scale
of order $\gevc{2}$ rather than \LQCD. Interpreted as a length (0.1~fm) it
suggests that, e.g., a red charge is ``neutralized'' by an anti-red charge at
this scale~\cite{Dischler:2000pk}. This is perhaps interesting as it suggests
that color is organized at a much smaller scale inside the hadrons and could
help explain how collective properties can arise in small systems.

With these final words I hope to stress my personal perspective that the high
quality data shown at this conference warrant theoretical ideas to be
developed for how to falsify the underlying assumptions.





\bibliographystyle{elsarticle-num} 
\bibliography{biblio}







\end{document}